\documentclass[english,pre,a4paper,twocolumn,noshowpacs,amsmath,amssymb]{revtex4-1}
\usepackage{graphicx,color,soul}

\newcommand{\dd}{\mathrm{d}}

\newcommand{\mean}[1]{\langle #1 \rangle}

\newcommand{\IInt}[3]{\int_{#2}^{#3}\dd #1\;}
\renewcommand{\vec}[1]{\mathbf #1}

\newcommand{\al}{\alpha}
\newcommand{\gam}{\gamma}
\newcommand{\eps}{\varepsilon}

\newcommand{\sig}{\sigma}
\newcommand{\om}{\omega}

\newcommand{\To}{T_\text{o}}
\newcommand{\Tm}{T_\text{m}}
\newcommand{\Tg}{T_\text{g}}
\newcommand{\TK}{T_\text{K}}
\newcommand{\df}{d_\text{f}}
\newcommand{\Ea}{E_\text{a}}
\newcommand{\Sc}{s_\text{c}}

\begin{document}

\title{Dynamic facilitation theory: A statistical mechanics approach to dynamic arrest}

\author{Thomas Speck}
\affiliation{Institut f\"ur Physik, Johannes Gutenberg-Universit\"at Mainz, Staudingerweg 7-9, 55128 Mainz, Germany}

\begin{abstract}
  The modeling of supercooled liquids approaching dynamic arrest has a long tradition, which is documented through a plethora of competing theoretical approaches. Here, we review the modeling of supercooled liquids in terms of dynamic ``defects'', also called excitations or soft spots, that are able to sustain motion. To this end, we consider a minimal statistical mechanics description in terms of active regions with the order parameter related to their typical size. This is the basis for both Adam-Gibbs and dynamical facilitation theory, which differ in their relaxation mechanism as the liquid is cooled: collective motion of more and more particles \emph{vs.} concerted hierarchical motion over larger and larger length scales. For the latter, dynamic arrest is possible without a growing static correlation length, and we sketch the derivation of a key result: the parabolic law for the structural relaxation time. We critically discuss claims in favor of a growing static length and argue that the resulting scenarios for pinning and dielectric relaxation are in fact compatible with dynamic facilitation.
\end{abstract}

\maketitle


\section{Introduction}

\begin{figure}[b!]
  \centering
  \includegraphics{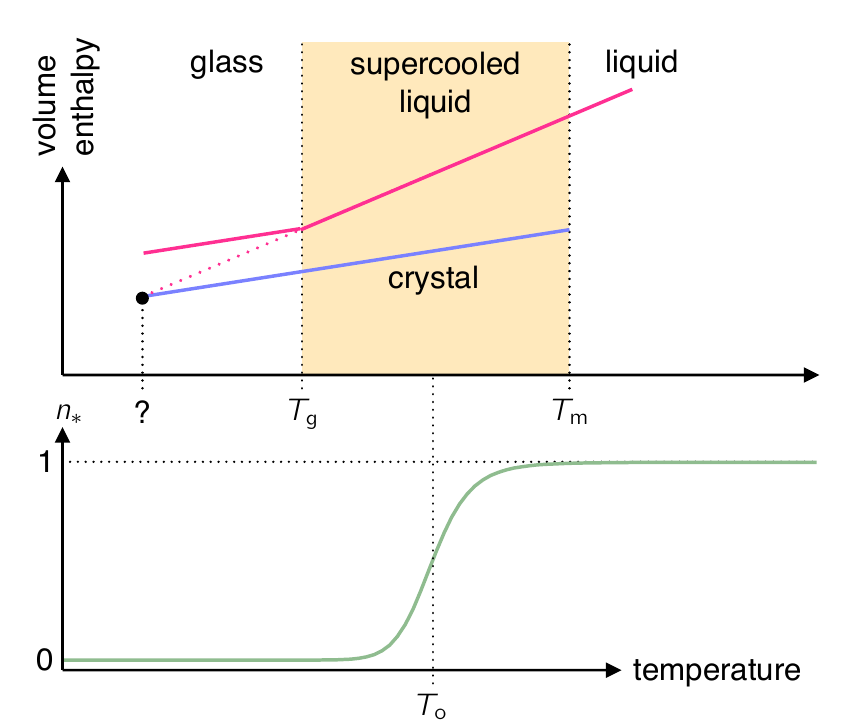}
  \caption{Top: Sketch of the evolution of molar volume or enthalpy as the liquid is cooled below the melting temperature $\Tm$. At $\Tg$, the liquid falls out of equilibrium and becomes a glass. This ``transition'' to a glass is to be understood as a dynamic crossover, not a thermodynamic transition. Naive extrapolation of the liquid to lower temperatures suggests an entropy crisis. Bottom: Population $n_\ast=Z_\ast/(1+Z_\ast)$ of excited states, which drops below the onset temperature $\To$. We normalize $Z_\ast(\To)=1$.}
  \label{fig:cool}
\end{figure}

Dynamic arrest is a generic phenomenon that occurs in a wide range of materials from bulk metallic glasses~\cite{wang04} to colloidal gels~\cite{zacc07,rich18}. Cooling a liquid below its melting temperature $\Tm$, the transformation to an ordered solid occurs through \emph{nucleation}. Accordingly, there is a time span before a sufficiently large nucleus appears, and during which the system remains a liquid (but supercooled). However, many liquids never crystallize but below some temperature $\Tg<\Tm$ become a non-equilibrium amorphous solid, a \emph{glass} (Fig.~\ref{fig:cool}). What is so puzzling is that the microscopic structure (as measured by the structure factor) remains that of the normal liquid while the dynamics (as measured by viscosity or structural relaxation time) slows down dramatically (increasing by more than ten orders of magnitude) on approaching the glass. Strongly supercooled liquids and glasses thus seem to defy the notion that structure determines the properties of a material.

Nevertheless, the currently dominating line of thought posits that the cause of the dynamic arrest somehow is related to a reduction of configurational entropy: there are less configurations in which the liquid's constituents (from now on identified with ``particles'') can be arranged, and exploring these configurations requires particles to move collectively. Support for this picture comes from exact results of mean-field models~\cite{pari10,char13a}. The question whether the mean-field scenario remains at least qualitatively valid in the presence of fluctuations (in three dimensions) is currently studied intensively~\cite{char17,scalliet17}.

Quite in contrast, kinetically constrained models~\cite{rito03,garr03,pan05,teomy14} demonstrate that dynamic arrest occurs even in systems with constant (or at least trivial) configurational entropy if dynamical rules are sufficiently complex. To understand how this is possible even for strictly local rules (the motion of a particle only depends on its current configuration), recall the Ising model of spins with nearest-neighbor interactions. Below its critical temperature, an ordered phase emerges in which spins are aligned with non-vanishing magnetization. Strikingly, the liquid-gas transition can be understood in exactly the same model. Close to the transition, both liquid and gas compete, leading to enhanced density fluctuations that, \emph{inter alia}, underlie the physical explanation of wetting and the hydrophobic effect~\cite{chan05a}. In supercooled liquids, something similar is observed: the short-time dynamics becomes strongly heterogeneous with large regions of the material being immobile while particles in small pockets remains highly mobile~\cite{dynhet11}. Is this \emph{dynamic heterogeneity} the signature of two competing \emph{dynamic} phases? This is the premise of dynamic facilitation theory~\cite{chan10}, which finds support from simulations of atomistic model glass formers~\cite{hedg09,turci18} and colloid experiments~\cite{pinc17}.

In atomistic glass formers, kinetic constraints are an \emph{emergent} phenomenon. While providing an explicit construction for the corresponding order parameter from particle positions is non-trivial, dynamic arrest follows from a few generic and plausible properties of this order parameter. In the first part of this manuscript, we review such a statistical modeling of ``mobility'', which is to be contrasted with theories such as model coupling theory~\cite{gotz92} that derive explicit dynamical equations. We then focus on dynamic facilitation: motion on smaller scales begets motion on larger scales. Our exposition is inspired by Keys \emph{et al.}~\cite{keys11} and does not explicitly reference kinetically constrained models, still leading to the same temperature dependence of the structural relaxation. In the second part, we apply this framework to several observations that have been argued to disfavor a dynamic view of the glass transition. Before starting, we emphasize that there are many excellent reviews~\cite{cava09} covering different theoretical aspects such as the landscape picture~\cite{debe01}, random-first order transition~\cite{lubc07}, and the role of local structure~\cite{roya15}.



\section{Statistical mechanics of mobility}

\subsection{Excited states}

The basic picture we have in mind is that the supercooled liquid can be divided into coarse-grained \emph{regions} (to be defined more precisely below) with qualitatively very different behaviors: Whereas most regions are jammed and particle motion is strongly hindered (whence inactive), some regions do allow for particle motion and are thus termed active. While this primarily seems to be a statement about dynamics, it also relates to structure in the sense that active regions are somehow ``softer'' in order to sustain motion.

Local particle motion within active regions, however, is not yet enough for structural relaxation since this motion is spatially confined and does not extend to the jammed regions. For relaxation to occur, we imagine that an active region has to transform into an \emph{excited state} that affects its neighboring regions. This transformation incurs a free energy cost. Assuming ergodicity, the ratio of average time $\tau_1$ spent in an active state and the lifetime $\tau_\ast$ of excited states is given by
\begin{equation}
  \label{eq:taual}
  \frac{\tau_1}{\tau_\ast} = \frac{Z_1}{Z_\ast}
\end{equation}
with corresponding partition functions $Z_1$ and $Z_\ast$. The structural relaxation time $\tau_\al\approx\tau_1$ is dominated by the waiting time to visit an excited state out of an active state. In a normal liquid, excited states are abundant with (apparent) activation energy $\Ea$, $Z_\ast=Z_1 e^{-\Ea/T}$. Throughout, we consider entropy to be dimensionless and measure temperature in units of energy. Both $\tau_\ast$ and $\Ea$ are assumed to be material properties and independent of temperature. The relaxation time then follows the Arrhenius law
\begin{equation}
  \label{eq:arr}
  \tau_\al(T) = \tau_\ast e^{\Ea/T}.
\end{equation}
Below an onset temperature $\To$, excited states become scarce and $Z_\ast$ drops rapidly with decreasing temperature, which is sketched in Fig.~\ref{fig:cool}. The behavior of the relaxation time then departs from the Arrhenius law Eq.~\eqref{eq:arr}.


\subsection{Adam-Gibbs theory}

\begin{figure}[b!]
  \centering
  \includegraphics{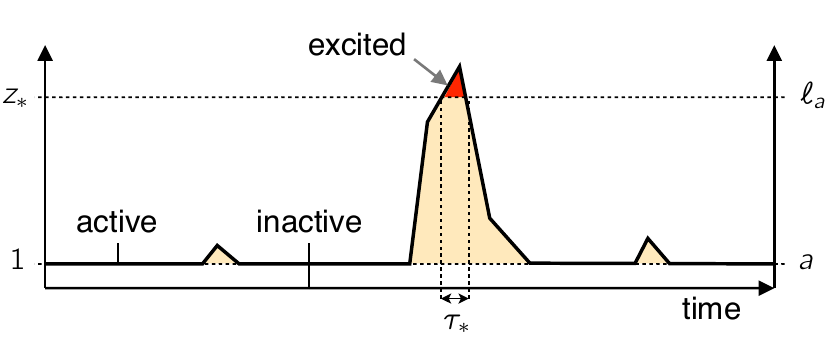}
  \caption{Time evolution of ``activity'' in a given region. On the left axis the size $z$ (in the sense of Adam-Gibbs) is indicated, while the right axis shows the extent $a$ over which motion is supported. Structural relaxation occurs when the region becomes excited by either reaching size $z^\ast$ or length $\ell_a$.}
  \label{fig:relax}
\end{figure}

Adam and Gibbs posit that structural relaxation is only possible if active regions have a certain minimal number $z\geq z_\ast$ of excited particles (cf. Fig.~\ref{fig:relax})~\cite{adam65}. The partition function for an active region composed of $z$ excited particles is assumed to be
\begin{equation}
  Z(z,T) = Z_1(T) e^{-z\Ea/T},
\end{equation}
where $\Ea$ is again the energy cost per excited particle. At low temperatures, the partition function $Z_\ast(T)=\sum_{z=z_\ast}^\infty Z(z,T)$ for a sufficiently large active region to be present in the liquid is dominated by the contribution of the smallest required size, $Z_\ast(T)\approx Z(z_\ast,T)$. To evaluate the critical size $z_\ast$, the crucial assumption is that the configurational entropy of an excited region needs to exceed a threshold $s_\ast$ corresponding to an ``excess'' of available configurations that make the reorganization possible. Hence, denoting $\Sc(T)$ the configurational entropy per particle, $z\Sc\geq s_\ast$ and thus $z_\ast=s_\ast/\Sc$. Rearranging Eq.~(\ref{eq:taual}) we obtain for the structural relaxation time
\begin{equation}
  \label{eq:ag}
  \tau_\al(T) = \frac{\tau_\ast Z_1}{Z_\ast} = \tau_\ast e^{z_\ast \Ea/T}
  = \tau_\ast\exp\left\{\frac{C}{T\Sc(T)}\right\}
\end{equation}
with constant $C=\Ea s_\ast$. For $z_\ast=1$ we recover the Arrhenius law Eq.~(\ref{eq:arr}). A super-Arrhenius increase of the relaxation time $\tau_\al$ is then obtained from a configurational entropy that drops as the temperature is lowered. Consequently, if $\Sc(\TK)\to0$ at a non-zero temperature $\TK$, this functional form would predict a divergence of the relaxation time.

This picture is refined into random-first order transition (RFOT) theory through introducing, in $d$ dimensions, an ``interfacial tension'' $\Upsilon z^{\theta/d}$ between amorphously ordered regions forming a ``mosaic''~\cite{kirk89,bouchaud04}. This implies a typical size of regions given by the (dimensionless) point-to-set length $\ell_\text{PS}=[\Upsilon/(T\Sc)]^\frac{1}{d-\theta}$. Structural relaxation is supposed to follow $\tau_\al=\tau_\ast\exp\{\Delta_0\ell_\text{PS}^\psi/T\}$ with another exponent $\psi$. The Adam-Gibbs form Eq.~\eqref{eq:ag} is recovered for $\psi/(d-\theta)=1$.


\subsection{Dynamical facilitation}
\label{sec:df}

\begin{figure}[t]
  \centering
  \includegraphics{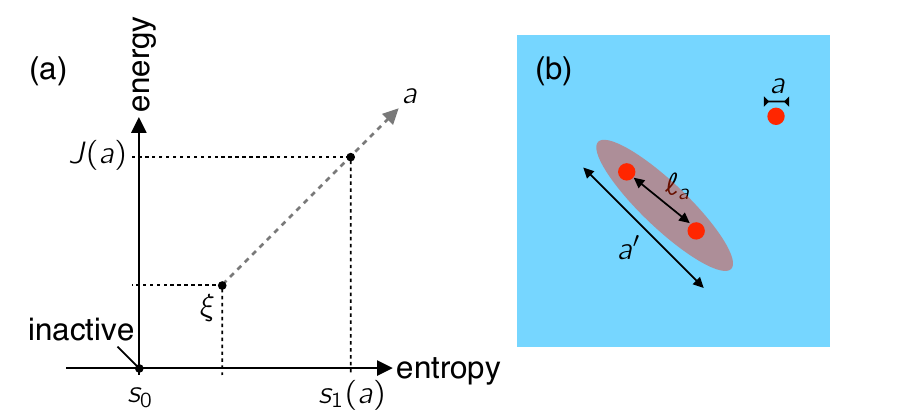}
  \caption{(a)~Energy \emph{vs.} entropy for active regions. As origin we choose the inactive state with zero energy and entropy $s_0$, which is independent of scale. Active states depend on the scale $a$ and lie on a line, the slope of which is given by the onset temperature $\To$. There is a lower bound $\xi$ below which mobility is indistinguishable from vibrations. (b)~Sketch of the relaxation mechanism. In order to relax two excitations of scale $a$ with typical separation $\ell_a$, a larger excitation of extent $a'\simeq\ell_a$ has to be borne out of thermal fluctuations.}
  \label{fig:sketch}
\end{figure}

Instead of using the size $z$, suppose we have an order parameter $a$ to characterize active regions. In particular, we assume that $a$ is a length that quantifies the linear extent over which an active region can support motion. For different $a$ these regions overlap, and for larger $a$ regions become sparser. For an active region there is a free energy cost $F_1(a,T)=J(a)-T\Delta s(a)$ modeling two effects: (i)~we need to excite particles with energetic cost $J(a)$ and (ii)~inactive regions, which we envision to be more rigid, have a lower entropy $s_0$ with $\Delta s(a)=s_1(a)-s_0$. Hence,
\begin{equation}
  \label{eq:Z:df}
  Z_1(a,T) = e^{-F_1/T} = e^{-J/(1/T-1/\To)},
\end{equation}
where we have defined the \emph{onset temperature}
\begin{equation}
  \label{eq:T:on}
  \To = \frac{J}{\Delta s}.
\end{equation}
For $T\gg\To$ active regions will be abundant while for $T\ll\To$ there will be very few. At low temperatures, the equilibrium concentration of active regions is
\begin{equation}
  \label{eq:c}
  c(a,T) \propto Z_1(a,T) = e^{-J(a)/\tilde T}, \qquad (T\ll\To)
\end{equation}
where we abbreviate $1/\tilde T=1/T-1/\To$. While the activation energy $J(a)$ and entropy difference $\Delta s(a)$ depend on the length $a$ (but not temperature), we will assume that $\To$ is a material property and thus independent of $a$ in the following. This implies a linear relationship of energy and entropy as shown in Fig.~\ref{fig:sketch}(a). Exactly at the onset temperature, the concentration $c$ is independent of $a$.

While active regions already support motion, as for Adam-Gibbs we assume that this is not enough for structural relaxation to occur. In contrast to a threshold size, however, we now posit that active regions have to ``connect'' to trigger the relaxation of the corresponding part of the liquid. While they could eventually achieve this through diffusion, let us further assume that the dominant mechanism is through the creation of ``larger'' active regions spanning the typical distance
\begin{equation}
  \label{eq:ell}
  \frac{\ell_a}{a} \sim [c(a,T)]^{-1/\df}
\end{equation}
between two smaller regions with length $a$ (we allow for non-compact regions through the fractal dimension $\df\leq d$). This mechanism is sketched in Fig.~\ref{fig:sketch}(b).

Again invoking Eq.~(\ref{eq:taual}) with $Z_\ast(a,T)=Z_1(\ell_a,T)$, the typical time for such a larger active region to be borne out of thermal fluctuations is
\begin{equation}
  \tau_1(a,T) = \frac{\tau_\ast Z_1}{Z_\ast} = \tau_\ast e^{\Delta F/T},
\end{equation}
which depends on the length scale $a$. The free energy difference $\Delta F=F_\ast-F_1$ between excited and active states reads
\begin{equation}
  \label{eq:barrier}
  \begin{split}
    \Delta F(a,T) &= [J(\ell_a) - T\Delta s(\ell_a)] - [J(a) - T\Delta s(a)] \\
    &= [J(\ell_a)-J(a)]\frac{T}{\tilde T},
  \end{split}
\end{equation}
where we have plugged in our definition~(\ref{eq:T:on}) for the onset temperature.

How does this larger excitation come about? In the facilitation picture~\cite{palm84}, motion on smaller scales begets motion on larger scales. To be more precise, we assume that there is a lower limit $a\geq\xi$ for which the construction of active regions can be carried out, \emph{i.e.}, for smaller lengths particles simply undergo ``wiggle'' motion from which the commitment to a new position cannot be discerned anymore~\cite{keys11}. One could think of these as ``elementary excitations''. To support a larger active region on $a=2\xi$ we need $n$ connected elementary excitations so that their density becomes $c(2\xi)=c(\xi)/n$. The next step of this hierarchy is $c(4\xi)=c(2\xi)/n=c(\xi)/n^2$ and so on. Hence, we can write the concentration as a power law
\begin{equation}
  c(a,T) = c(\xi,T)(a/\xi)^{-\al(T)}
\end{equation}
with $n=2^\al$, where the exponent $\al(T)$ depends on temperature. An obvious consequence of this scale-invariant form is that for a different length $a'$ we obtain
\begin{equation}
  \label{eq:c:sf}
  c(a',T) = c(a,T)(a'/a)^{-\al(T)}
\end{equation}
independent of $\xi$. 

We have now two relations for the dependence of the concentration on scale $a$. Plugging Eq.~(\ref{eq:c}) for temperatures below the onset temperature into Eq.~(\ref{eq:c:sf}) we obtain the prediction
\begin{equation}
  \label{eq:J}
  J(a) - J(a') = \al\tilde T\ln(a/a') = \gam J(\sig)\ln(a/a')
\end{equation}
for the energy cost of active regions. To be consistent with temperature-independent energies $J(a)$, the prefactor $\al(T)\tilde T$ needs to be independent of temperature. In the second step, we have rewritten this prefactor using as reference the energy $J(\sig)$ together with a dimensionless factor $\gam$. This reference is chosen to correspond to the length scale on which the structural relaxation time $\tau_\al$ is probed, which typically corresponds to the particle diameter $\sig$. Plugging Eq.~(\ref{eq:J}) into Eq.~(\ref{eq:barrier}) and using Eq.~(\ref{eq:c}) with $a=\sig$, we finally obtain the ``parabolic law''
\begin{equation}
  \tau_\al(T) = \tau_\ast\exp\left\{\frac{\mathcal J^2}{\To^2}\left(\frac{\To}{T}-1\right)^2\right\}
\end{equation}
for the structural relaxation time with effective energy scale $\mathcal J=J(\sig)\sqrt{\gam/\df}$. This functional form has been shown to describe very well the relaxation times (and viscosities) of a wide range of glass formers~\cite{elma09}. In contrast to Eq.~\eqref{eq:ag}, there is no singularity and $\tau_\al$ diverges only as $T\to0$.


\section{Discussion}

\subsection{Constructing the order parameter}

So far, we have explored the statistical consequences of active and excited regions through which relaxation occurs. We have assumed that these regions can be characterized through a length $a$, which should be related to their linear extend. Clearly, this is not a very precise definition and confronted with configurations of particle positions generated from computer simulations (or experimentally accessible data), one wonders how to construct a suitable order parameter.

One route is to identify active regions from the \emph{short-time} motion of single particles. Typically, a particle is considered ``mobile'' or ``rearranging'' if it has moved at least the distance $a$ over a certain time, which is chosen to be large enough to allow particles to rearrange but small enough to prevent detecting multiple jumps as one event. The rationale is that mobile particles indicate the presence of an active region, and the partition function $Z_1$ is then estimated from the fraction of mobile particles. The numerical results presented in Ref.~\citenum{keys11} for various model glass formers support the form assumed in Eq.~(\ref{eq:Z:df}) for the concentration of excitations.

A second route is to correlate dynamics with local structure. There are different descriptors for local motifs beyond pair correlations, \emph{e.g.}, common neighbor analysis~\cite{hone87}, Voronoi polyhedra~\cite{cosl11}, and the topological cluster classification (TCC)~\cite{mali13}. Promoting certain motifs has been shown to induce dynamic arrest in model glass formers~\cite{spec12b,turci17,turci18}. However, a specific (model-dependent) local structural motif that unambiguously correlates with single particle motion has remained elusive~\cite{hock14,cosl16}.

Yet another, more ``agnostic'' strategy is followed in Ref.~\citenum{scho16}, where a large pool of structural measures is used to describe the local environment of each particle. Molecular dynamics simulations are used to train a support vector machine to identify ``soft'' particles that will rearrange based on their local environment. The distance to the hyperplane separating soft from hard particles then defines the softness $S$. The probability for a particle to rearrange is shown to follow
\begin{equation}
  P_1(S) = e^{-\Delta E(S)/T+\Sigma(S)}
\end{equation}
with temperature-independent functions $\Delta E(S)$ and $\Sigma(S)$. Since $P_1\propto Z_1$ we recover Eq.~\eqref{eq:Z:df}, where $S$ replaces $a$, and $\Delta E$ is related to $J$ and $\Sigma$ to the entropy change $\Delta s$ (up to additive constants due to the normalization). Hence, the softness $S$ provides a concrete construction for an order parameter that reflects the statistical properties we had assumed.
%

\subsection{Configurational entropy}

In Fig.~\ref{fig:cool}, the dependency of enthalpy on temperature is sketched. Assuming that the liquid remains in equilibrium, extrapolating the shown behavior to lower temperatures seems to cross the crystal at a \emph{non-zero} temperature. This observation is called the ``Kauzmann'' paradox~\cite{kauz48} and is often cited as argument for a thermodynamic phase transition into an ``ideal glass'' with only subextensive configurational entropy. However, the presence of active regions removes this singularity, which follows from a simple argument by Stillinger~\cite{stil88}: Consider a putative ``ideal glass'' configurations of $N$ particles with inherent state energy per particle $\phi_0$. Now recall that a single elementary excitation costs a finite energy $\eps=J(\xi)$. For $n$ such non-interacting excitations, the configurational energy per particle becomes $\phi=\phi_0+\eps n/N$. Assuming that excitations correspond to structural ``vacancies'' (extra space to sustain motion), the number of different packings is $W(n)=(N+n)!/(N!n!)$, \emph{i.e.}, the number of combinations to place $N$ identical particles onto $N+n$ sites. Employing Stirling's formula and eliminating $n$ in favor of $\Delta\phi=\phi-\phi_0$, the configurational entropy $N\Sc=\ln W$ to leading order becomes
\begin{equation}
  \Sc(\phi) = -\frac{\Delta\phi}{\eps}\ln\frac{\Delta\phi}{\eps} + \mathcal O(\Delta\phi).
\end{equation}
The consequence of the logarithmic dependence is that the slope of the configurational entropy becomes infinite as $\phi\to\phi_0$. Since the derivative of the entropy with respect to energy yields the inverse temperature, the ideal glass could only be reached as $T\to0$.

\subsection{Heat capacity}

\begin{figure}[b!]
  \centering
  \includegraphics{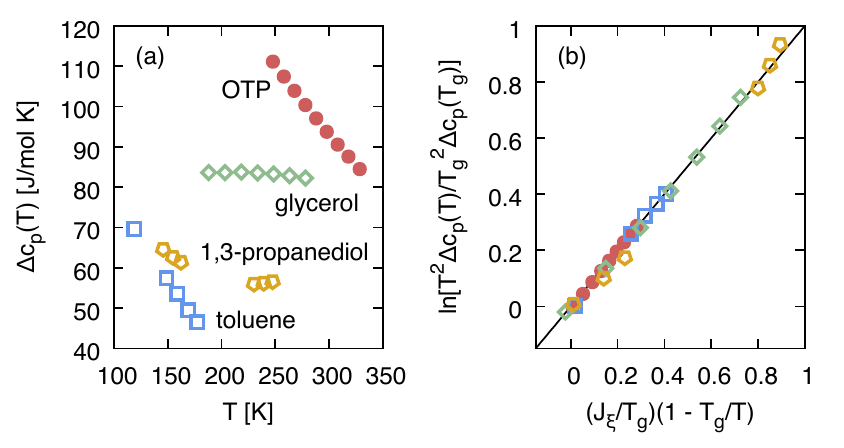}
  \caption{(a)~Experimental excess heat capacities $\Delta c_p$ for several molecular liquids approaching the glass transition temperature $\Tg$ (data from Ref.~\citenum{moyn00}). (b)~Collapse of data when plotted according to Eq.~\eqref{eq:heatcap}.}
  \label{fig:heatcap}
\end{figure}

We now turn to the excess heat capacity due to the presence of active regions, the temperature-dependence of which has caused some controversy previously~\cite{biro05,chan05}. Active regions have a higher energy so that the excess energy of the liquid compared to a completely jammed solid reads
\begin{equation}
  \Delta E(T) \sim \IInt{a}{\xi}{\infty} \frac{V}{a^d} J(a)c(a,T).
\end{equation}
At least close to the glass transition, the excess energy is dominated by the most-likely contribution from the elementary excitations with length $\xi$,
\begin{equation}
  \Delta E(T) \sim \frac{V}{\xi^d}J(\xi) e^{-J(\xi)/\tilde T}.
\end{equation}
At constant volume, the excess heat capacity is obtained as $\Delta c(T)=\partial(\Delta E)/\partial T$. Normalized by its value at $\Tg$, we obtain
\begin{equation}
  \label{eq:heatcap}
  \frac{\Delta c(T)}{\Delta c(\Tg)} = \frac{\Tg^2}{T^2}\exp\left\{ -\frac{J(\xi)}{\Tg}\left(\frac{\Tg}{T}-1\right) \right\}
\end{equation}
independent of the onset temperature. This expression should also hold at constant pressure. In Fig.~\ref{fig:heatcap}(a), four different datasets for molecular liquids obtained in Ref.~\citenum{moyn00} are plotted. While showing qualitatively different behavior, all datasets collapse when $\Delta c_p(T)/\Delta c_p(\Tg)$ is plotted as a function of rescaled temperature according to Eq.~\eqref{eq:heatcap}, cf. Fig.~\ref{fig:heatcap}(b). The only free parameter here is the ratio $J(\xi)/\Tg$, which is listed in Table~\ref{tab:heatcap} and for which we find values between $1$ and $2.5$. On physical grounds, one would indeed expect an activation energy $J(\xi)\sim\Tg$ that is on the order of the thermal energy. Moreover, if available we have estimated the energies $J(\sigma)$ from the fitted $\mathcal J$ provided in Ref.~\citenum{elma09}. In Ref.~\citenum{keys11} it is shown that $\gam\approx0.5$ and $\df\approx2.4$ for several simulated model glass formers, and we have used these values. Exploiting Eq.~\eqref{eq:J}, we can thus estimate the ratio $\xi/\sig$, which again gives reasonable numbers.

\begin{table}[t]
  \begin{tabular}{l|ccccc}
    \hline\hline
    liquid & $\Tg$ [K] & $J(\xi)/\Tg$ & $J(\sig)/\To$ & $\To$ [K] & $\xi/\sig$ \\
    \hline
    OTP & 246 & 1.13 & 16.9 & 357 & 0.15 \\
    toluene & 117 & 1.19 \\
    glycerol & 190 & 2.29 & 9.0 & 338 & 0.18 \\
    1,3-propanediol & 145 & 2.16 \\
    \hline\hline
  \end{tabular}
  \caption{For the datasets shown in Fig.~\ref{fig:heatcap} we list the dynamic glass transition temperatures $\Tg$. The second column lists the values of $J(\xi)$ obtained from fitting Eq.~\eqref{eq:heatcap} to the experimental heat capacities. For OTP and glycerol, we use results of Ref.~\citenum{elma09} for $\mathcal J$ and the onset temperature $\To$ to estimate $J(\sig)$ and the ratios $\xi/\sig$.}
  \label{tab:heatcap}
\end{table}

\subsection{Pinning}

A fundamental and conceptual problem is that the actual glass transition at $\Tg$ is (and most likely will be) out of reach of direct computer simulations of model glass formers. One strategy to address this issue has been to ``pin'' (immobilize) a fraction of randomly selected particles in an equilibrated liquid~\cite{camm12,gokh14}. Increasing the concentration $c_\text{p}$ of pinned particles, the liquid becomes more sluggish and finally reaches a glass state even at accessible temperatures. The transition line is predicted to emanate from the Kauzmann transition and to end in a critical point.

\begin{figure}[t]
  \centering
  \includegraphics{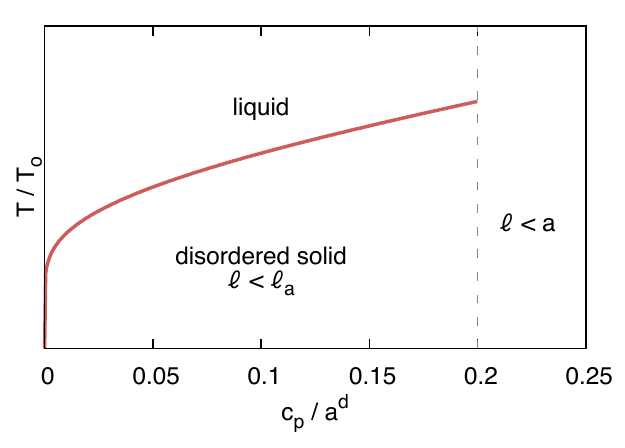}
  \caption{Sketch of the ``pinning phase diagram'' for the crossover from the ergodic liquid to a disordered arrested solid due to pinning a concentration $c_\text{p}$ of particles setting the length $\ell\sim c_\text{p}^{-1/d}$. The lines depend on the scale $a$. Below the solid line, relaxation through larger excitations (cf. Fig.~\ref{fig:sketch}) is suppressed. Beyond the dashed line, pinned particles even prevent active regions.}
  \label{fig:pinning}
\end{figure}

The qualitative picture obtained from dynamic facilitation is similar. The pinned particles induce a static length scale $\ell\sim c_\text{p}^{-1/d}$. They effectively suppress active regions with length scale larger than $\ell$. The physical picture is that, since the liquid is so dense, even the presence of a single pinned particle will destroy the delicate structural arrangement that allows for motion. Hence, for $\ell_a\gtrsim\ell$, relaxation cannot proceed anymore through the hierarchical pathway described in Sec.~\ref{sec:df} and, in the absence of other relaxation mechanisms, the material would be frozen into a single configuration. To estimate the temperature $T_\text{p}$ for this crossover, we solve $\ell_a\sim\ell$ using Eq.~\eqref{eq:ell},
\begin{equation}
  \frac{T_\text{p}(a)}{\To} \sim \left(1 - \frac{\To\df}{J(a)d}\ln\frac{c_\text{p}}{a^d}\right)^{-1}.
\end{equation}
This temperature depends on the scale $a$ and goes to zero as $c_\text{p}\to0$. We assume that for $\ell>a$ the entropy $s_1(a)$ of local mobile regions is not affected by the pinned particles such that the onset temperature $\To$ remains unchanged. For $\ell<a$ this assumption breaks down since any region of extend $a$ most likely contains pinned particles and thus cannot be active anymore. The resulting qualitative behavior is shown in Fig.~\ref{fig:pinning}, which strongly resembles the phase diagram of Ref.~\citenum{camm12} based on mean-field calculations. At variance, the dynamic facilitation picture yields crossovers instead of genuine transitions and, moreover, $T_\text{p}\to0$ as the concentration $c_\text{p}\to0$ of pinned particles vanishes.

\subsection{Dielectric relaxation}

High-precision measurements of the dielectric susceptibility $\chi(\om)$ of two molecular liquids~\cite{albe16} indicate that the normalized peaks of the non-linear susceptibilities $X_3^\ast$ and $X_5^\ast$ scale as $X_5^\ast\sim (X_3^\ast)^2$, which has been interpreted within the framework of RFOT in favor of a growing static length-scale. At variance with that interpretation, the argument we just used for pinning can be extended to describing dielectric relaxation. Now the external length scale $\ell$ is set by the condition $\om\tau_1(\ell)\sim 1$, with $\om$ the frequency of the applied electric field $E$. On smaller scales $a$ with $\ell_a<\ell$, the dynamics can follow the external field and we expect the response of a liquid. On the other hand, inhomogeneities on larger scales with $\ell_a>\ell$ do not relax on timescales $\om^{-1}$ and thus contribute to the glassy dielectric response.

The polarization density of an active region with extent $a$ is $P(a)=\mean{\hat\mu}/a^d$. Here, $\hat{\boldsymbol\mu}=\sum_{i=1}^n\mu_0\vec e_i$ is the dipole moment of the region, which we write as sum over $n\sim a^d$ elementary dipole moments $\mu_0$ with unit orientations $\vec e_i$. Performing the thermal average over a single super-dipole moment with maximal magnitude $\mu$ leads to
\begin{equation}
  \mean{\hat\mu} = \frac{\IInt{\hat\mu}{0}{\mu}\hat\mu^d \cdots}{\IInt{\hat\mu}{0}{\mu}\mu^{d-1} \cdots} = \mu F(\mu E/T)
\end{equation}
using spherical coordinates and assuming that the ellipses only involve the combination $\hat\mu E/T$. The undetermined scaling function $F(x)$ obeys $F(-x)=-F(x)$, cf. Ref.~\citenum{albe16}, which guarantees a vanishing polarization at zero external field.

Since motion is possible within active regions, we assume that they behave as super-dipoles but with uncorrelated elementary dipoles, $\mean{\vec e_i\cdot\vec e_j}=\delta_{ij}$, whereas inactive regions cannot be polarized. The maximal dipole moment then becomes $\mu=|\mean{\hat{\boldsymbol\mu}}|=\mu_0\sqrt{n}$. The average
\begin{equation}
  \mathcal P \sim \IInt{a}{\ell}{\infty} P(a)c(a,T) \sim P(\ell)c(\ell,T)
\end{equation}
over regions is again dominated by the smallest contributing length $\ell$. Expanding the polarization density in powers of $E$, $\mathcal P=\sum_k\chi_k E^k$, we obtain the peak susceptibilities at frequency $\om^\ast\sim1/\tau_\al$
\begin{equation}
  |\chi_k^\ast| \sim \frac{\mu_0^{k+1}}{T^k} c(\ell,T) \ell^\frac{(k-1)d}{2}
\end{equation}
dropping prefactors. For even $k$, $\chi_k=0$ due to the antisymmetry of $F(x)$. Following Ref.~\citenum{albe16}, we introduce the scaled susceptibilities
\begin{equation}
  X_3^\ast = \frac{T|\chi_3^\ast|}{|\chi_1^\ast|^2} \sim \frac{\ell^d}{c(\ell,T)}
\end{equation}
and
\begin{equation*}
  X_5^\ast = \frac{T^2|\chi_5^\ast|}{|\chi_1^\ast|^3} \sim \frac{\ell^{2d}}{[c(\ell,T)]^2} \sim (X_3^\ast)^2.
\end{equation*}
Hence, we recover exactly the same relation $X_5^\ast\sim (X_3^\ast)^2$ that is found in Ref.~\citenum{albe16}.


\section{Conclusions}

To summarize, we have given a brief introduction to the modeling of dynamic arrest through a statistical mechanics of mobility. We have contrasted Adam-Gibbs (AG) and dynamic facilitation (DF), which both start from a similar physical picture of localized active regions that sustain motion in a sea of immobility. The main pathway for structural relaxation, however, differs: AG posits that particles have to move collectively while DF posits that a larger active region has to be born out of motion on smaller scales. The most striking difference is that AG predicts a diverging relaxation time at a finite temperature while such a singularity is absent in DF. Over the range of accessible temperatures, however, both yield acceptable fits to the data. To discriminate these theoretical approaches, one thus needs to turn to other observables. DF has been shown to account successfully for the breakdown of the Stokes-Einstein relation~\cite{jung04,pan05} and the cooling-rate dependence of heat capacities from differential scanning calorimetry~\cite{keys13a}. Two further candidates discussed here are the response to pinning a fraction of particles and dielectric relaxation. While both have been claimed to probe a growing \emph{static} length scale, we have argued that they are compatible with a growing \emph{dynamic} length scale.


\acknowledgments

I dedicate this manuscript to the memory of David Chandler. I am deeply thankful to Rob Jack, Paddy Royall, and Juan Garrahan for countless inspiring discussions on the nature of dynamic arrest.


%
  

\begin{thebibliography}{46}%
  \makeatletter
  \providecommand \@ifxundefined [1]{%
   \@ifx{#1\undefined}
  }%
  \providecommand \@ifnum [1]{%
   \ifnum #1\expandafter \@firstoftwo
   \else \expandafter \@secondoftwo
   \fi
  }%
  \providecommand \@ifx [1]{%
   \ifx #1\expandafter \@firstoftwo
   \else \expandafter \@secondoftwo
   \fi
  }%
  \providecommand \natexlab [1]{#1}%
  \providecommand \enquote  [1]{``#1''}%
  \providecommand \bibnamefont  [1]{#1}%
  \providecommand \bibfnamefont [1]{#1}%
  \providecommand \citenamefont [1]{#1}%
  \providecommand \href@noop [0]{\@secondoftwo}%
  \providecommand \href [0]{\begingroup \@sanitize@url \@href}%
  \providecommand \@href[1]{\@@startlink{#1}\@@href}%
  \providecommand \@@href[1]{\endgroup#1\@@endlink}%
  \providecommand \@sanitize@url [0]{\catcode `\\12\catcode `\$12\catcode
    `\&12\catcode `\#12\catcode `\^12\catcode `\_12\catcode `\%12\relax}%
  \providecommand \@@startlink[1]{}%
  \providecommand \@@endlink[0]{}%
  \providecommand \url  [0]{\begingroup\@sanitize@url \@url }%
  \providecommand \@url [1]{\endgroup\@href {#1}{\urlprefix }}%
  \providecommand \urlprefix  [0]{URL }%
  \providecommand \Eprint [0]{\href }%
  \providecommand \doibase [0]{http://dx.doi.org/}%
  \providecommand \selectlanguage [0]{\@gobble}%
  \providecommand \bibinfo  [0]{\@secondoftwo}%
  \providecommand \bibfield  [0]{\@secondoftwo}%
  \providecommand \translation [1]{[#1]}%
  \providecommand \BibitemOpen [0]{}%
  \providecommand \bibitemStop [0]{}%
  \providecommand \bibitemNoStop [0]{.\EOS\space}%
  \providecommand \EOS [0]{\spacefactor3000\relax}%
  \providecommand \BibitemShut  [1]{\csname bibitem#1\endcsname}%
  \let\auto@bib@innerbib\@empty
  \bibitem [{\citenamefont {Wang}\ \emph {et~al.}(2004)\citenamefont {Wang},
    \citenamefont {Dong},\ and\ \citenamefont {Shek}}]{wang04}%
    \BibitemOpen
    \bibfield  {author} {\bibinfo {author} {\bibfnamefont {W.}~\bibnamefont
    {Wang}}, \bibinfo {author} {\bibfnamefont {C.}~\bibnamefont {Dong}}, \ and\
    \bibinfo {author} {\bibfnamefont {C.}~\bibnamefont {Shek}},\ }\href {\doibase
    10.1016/j.mser.2004.03.001} {\bibfield  {journal} {\bibinfo  {journal}
    {Mater. Sci. Eng. R Rep.}\ }\textbf {\bibinfo {volume} {44}},\ \bibinfo
    {pages} {45} (\bibinfo {year} {2004})}\BibitemShut {NoStop}%
  \bibitem [{\citenamefont {Zaccarelli}(2007)}]{zacc07}%
    \BibitemOpen
    \bibfield  {author} {\bibinfo {author} {\bibfnamefont {E.}~\bibnamefont
    {Zaccarelli}},\ }\href {\doibase 10.1088/0953-8984/19/32/323101} {\bibfield
    {journal} {\bibinfo  {journal} {J. Phys.: Condens. Matter}\ }\textbf
    {\bibinfo {volume} {19}},\ \bibinfo {pages} {323101} (\bibinfo {year}
    {2007})}\BibitemShut {NoStop}%
  \bibitem [{\citenamefont {Richard}\ \emph {et~al.}(2018)\citenamefont
    {Richard}, \citenamefont {Hallett}, \citenamefont {Speck},\ and\
    \citenamefont {Royall}}]{rich18}%
    \BibitemOpen
    \bibfield  {author} {\bibinfo {author} {\bibfnamefont {D.}~\bibnamefont
    {Richard}}, \bibinfo {author} {\bibfnamefont {J.}~\bibnamefont {Hallett}},
    \bibinfo {author} {\bibfnamefont {T.}~\bibnamefont {Speck}}, \ and\ \bibinfo
    {author} {\bibfnamefont {C.~P.}\ \bibnamefont {Royall}},\ }\href {\doibase
    10.1039/c8sm00389k} {\bibfield  {journal} {\bibinfo  {journal} {Soft Matter}\
    }\textbf {\bibinfo {volume} {14}},\ \bibinfo {pages} {5554} (\bibinfo {year}
    {2018})}\BibitemShut {NoStop}%
  \bibitem [{\citenamefont {Parisi}\ and\ \citenamefont
    {Zamponi}(2010)}]{pari10}%
    \BibitemOpen
    \bibfield  {author} {\bibinfo {author} {\bibfnamefont {G.}~\bibnamefont
    {Parisi}}\ and\ \bibinfo {author} {\bibfnamefont {F.}~\bibnamefont
    {Zamponi}},\ }\href {\doibase 10.1103/RevModPhys.82.789} {\bibfield
    {journal} {\bibinfo  {journal} {Rev. Mod. Phys.}\ }\textbf {\bibinfo {volume}
    {82}},\ \bibinfo {pages} {789} (\bibinfo {year} {2010})}\BibitemShut
    {NoStop}%
  \bibitem [{\citenamefont {Charbonneau}\ \emph {et~al.}(2014)\citenamefont
    {Charbonneau}, \citenamefont {Kurchan}, \citenamefont {Parisi}, \citenamefont
    {Urbani},\ and\ \citenamefont {Zamponi}}]{char13a}%
    \BibitemOpen
    \bibfield  {author} {\bibinfo {author} {\bibfnamefont {P.}~\bibnamefont
    {Charbonneau}}, \bibinfo {author} {\bibfnamefont {J.}~\bibnamefont
    {Kurchan}}, \bibinfo {author} {\bibfnamefont {G.}~\bibnamefont {Parisi}},
    \bibinfo {author} {\bibfnamefont {P.}~\bibnamefont {Urbani}}, \ and\ \bibinfo
    {author} {\bibfnamefont {F.}~\bibnamefont {Zamponi}},\ }\href {\doibase
    10.1038/ncomms4725} {\bibfield  {journal} {\bibinfo  {journal} {Nat.
    Commun.}\ }\textbf {\bibinfo {volume} {5}},\  (\bibinfo {year}
    {2014})}\BibitemShut {NoStop}%
  \bibitem [{\citenamefont {Charbonneau}\ \emph {et~al.}(2017)\citenamefont
    {Charbonneau}, \citenamefont {Kurchan}, \citenamefont {Parisi}, \citenamefont
    {Urbani},\ and\ \citenamefont {Zamponi}}]{char17}%
    \BibitemOpen
    \bibfield  {author} {\bibinfo {author} {\bibfnamefont {P.}~\bibnamefont
    {Charbonneau}}, \bibinfo {author} {\bibfnamefont {J.}~\bibnamefont
    {Kurchan}}, \bibinfo {author} {\bibfnamefont {G.}~\bibnamefont {Parisi}},
    \bibinfo {author} {\bibfnamefont {P.}~\bibnamefont {Urbani}}, \ and\ \bibinfo
    {author} {\bibfnamefont {F.}~\bibnamefont {Zamponi}},\ }\href {\doibase
    10.1146/annurev-conmatphys-031016-025334} {\bibfield  {journal} {\bibinfo
    {journal} {Annu. Rev. Condens. Matter Phys.}\ }\textbf {\bibinfo {volume}
    {8}},\ \bibinfo {pages} {265} (\bibinfo {year} {2017})}\BibitemShut {NoStop}%
  \bibitem [{\citenamefont {Scalliet}\ \emph {et~al.}(2017)\citenamefont
    {Scalliet}, \citenamefont {Berthier},\ and\ \citenamefont
    {Zamponi}}]{scalliet17}%
    \BibitemOpen
    \bibfield  {author} {\bibinfo {author} {\bibfnamefont {C.}~\bibnamefont
    {Scalliet}}, \bibinfo {author} {\bibfnamefont {L.}~\bibnamefont {Berthier}},
    \ and\ \bibinfo {author} {\bibfnamefont {F.}~\bibnamefont {Zamponi}},\ }\href
    {\doibase 10.1103/PhysRevLett.119.205501} {\bibfield  {journal} {\bibinfo
    {journal} {Phys. Rev. Lett.}\ }\textbf {\bibinfo {volume} {119}},\ \bibinfo
    {pages} {205501} (\bibinfo {year} {2017})}\BibitemShut {NoStop}%
  \bibitem [{\citenamefont {Ritort}\ and\ \citenamefont
    {Sollich}(2003)}]{rito03}%
    \BibitemOpen
    \bibfield  {author} {\bibinfo {author} {\bibfnamefont {F.}~\bibnamefont
    {Ritort}}\ and\ \bibinfo {author} {\bibfnamefont {P.}~\bibnamefont
    {Sollich}},\ }\href {\doibase 10.1080/0001873031000093582} {\bibfield
    {journal} {\bibinfo  {journal} {Adv. Phys.}\ }\textbf {\bibinfo {volume}
    {52}},\ \bibinfo {pages} {219} (\bibinfo {year} {2003})}\BibitemShut
    {NoStop}%
  \bibitem [{\citenamefont {Garrahan}\ and\ \citenamefont
    {Chandler}(2003)}]{garr03}%
    \BibitemOpen
    \bibfield  {author} {\bibinfo {author} {\bibfnamefont {J.~P.}\ \bibnamefont
    {Garrahan}}\ and\ \bibinfo {author} {\bibfnamefont {D.}~\bibnamefont
    {Chandler}},\ }\href {\doibase 10.1073/pnas.1233719100} {\bibfield  {journal}
    {\bibinfo  {journal} {Proc. Natl. Acad. Sci. U.S.A.}\ }\textbf {\bibinfo
    {volume} {100}},\ \bibinfo {pages} {9710} (\bibinfo {year}
    {2003})}\BibitemShut {NoStop}%
  \bibitem [{\citenamefont {Pan}\ \emph {et~al.}(2005)\citenamefont {Pan},
    \citenamefont {Garrahan},\ and\ \citenamefont {Chandler}}]{pan05}%
    \BibitemOpen
    \bibfield  {author} {\bibinfo {author} {\bibfnamefont {A.~C.}\ \bibnamefont
    {Pan}}, \bibinfo {author} {\bibfnamefont {J.~P.}\ \bibnamefont {Garrahan}}, \
    and\ \bibinfo {author} {\bibfnamefont {D.}~\bibnamefont {Chandler}},\ }\href
    {\doibase 10.1103/PhysRevE.72.041106} {\bibfield  {journal} {\bibinfo
    {journal} {Phys. Rev. E}\ }\textbf {\bibinfo {volume} {72}},\ \bibinfo
    {pages} {041106} (\bibinfo {year} {2005})}\BibitemShut {NoStop}%
  \bibitem [{\citenamefont {Teomy}\ and\ \citenamefont {Shokef}(2014)}]{teomy14}%
    \BibitemOpen
    \bibfield  {author} {\bibinfo {author} {\bibfnamefont {E.}~\bibnamefont
    {Teomy}}\ and\ \bibinfo {author} {\bibfnamefont {Y.}~\bibnamefont {Shokef}},\
    }\href {\doibase 10.1103/physreve.89.032204} {\bibfield  {journal} {\bibinfo
    {journal} {Phys. Rev. E}\ }\textbf {\bibinfo {volume} {89}},\ \bibinfo
    {pages} {032204} (\bibinfo {year} {2014})}\BibitemShut {NoStop}%
  \bibitem [{\citenamefont {Chandler}(2005)}]{chan05a}%
    \BibitemOpen
    \bibfield  {author} {\bibinfo {author} {\bibfnamefont {D.}~\bibnamefont
    {Chandler}},\ }\href {\doibase 10.1038/nature04162} {\bibfield  {journal}
    {\bibinfo  {journal} {Nature}\ }\textbf {\bibinfo {volume} {437}},\ \bibinfo
    {pages} {640} (\bibinfo {year} {2005})}\BibitemShut {NoStop}%
  \bibitem [{\citenamefont {Berthier}\ \emph {et~al.}(2011)\citenamefont
    {Berthier}, \citenamefont {Biroli}, \citenamefont {Bouchaud}, \citenamefont
    {Cipelletti},\ and\ \citenamefont {van Saarloos}}]{dynhet11}%
    \BibitemOpen
    \bibinfo {editor} {\bibfnamefont {L.}~\bibnamefont {Berthier}}, \bibinfo
    {editor} {\bibfnamefont {G.}~\bibnamefont {Biroli}}, \bibinfo {editor}
    {\bibfnamefont {J.-P.}\ \bibnamefont {Bouchaud}}, \bibinfo {editor}
    {\bibfnamefont {L.}~\bibnamefont {Cipelletti}}, \ and\ \bibinfo {editor}
    {\bibfnamefont {W.}~\bibnamefont {van Saarloos}},\ eds.,\ \href {\doibase
    10.1093/acprof:oso/9780199691470.001.0001} {\emph {\bibinfo {title}
    {Dynamical Heterogeneities in Glasses, Colloids, and Granular Media}}}\
    (\bibinfo  {publisher} {Oxford University Press},\ \bibinfo {year}
    {2011})\BibitemShut {NoStop}%
  \bibitem [{\citenamefont {Chandler}\ and\ \citenamefont
    {Garrahan}(2010)}]{chan10}%
    \BibitemOpen
    \bibfield  {author} {\bibinfo {author} {\bibfnamefont {D.}~\bibnamefont
    {Chandler}}\ and\ \bibinfo {author} {\bibfnamefont {J.~P.}\ \bibnamefont
    {Garrahan}},\ }\href@noop {} {\bibfield  {journal} {\bibinfo  {journal}
    {Annu. Rev. Phys. Chem.}\ }\textbf {\bibinfo {volume} {61}},\ \bibinfo
    {pages} {191} (\bibinfo {year} {2010})}\BibitemShut {NoStop}%
  \bibitem [{\citenamefont {Hedges}\ \emph {et~al.}(2009)\citenamefont {Hedges},
    \citenamefont {Jack}, \citenamefont {Garrahan},\ and\ \citenamefont
    {Chandler}}]{hedg09}%
    \BibitemOpen
    \bibfield  {author} {\bibinfo {author} {\bibfnamefont {L.~O.}\ \bibnamefont
    {Hedges}}, \bibinfo {author} {\bibfnamefont {R.~L.}\ \bibnamefont {Jack}},
    \bibinfo {author} {\bibfnamefont {J.~P.}\ \bibnamefont {Garrahan}}, \ and\
    \bibinfo {author} {\bibfnamefont {D.}~\bibnamefont {Chandler}},\ }\href
    {\doibase 10.1126/science.1166665} {\bibfield  {journal} {\bibinfo  {journal}
    {Science}\ }\textbf {\bibinfo {volume} {323}},\ \bibinfo {pages} {1309}
    (\bibinfo {year} {2009})}\BibitemShut {NoStop}%
  \bibitem [{\citenamefont {Turci}\ \emph {et~al.}(2018)\citenamefont {Turci},
    \citenamefont {Speck},\ and\ \citenamefont {Royall}}]{turci18}%
    \BibitemOpen
    \bibfield  {author} {\bibinfo {author} {\bibfnamefont {F.}~\bibnamefont
    {Turci}}, \bibinfo {author} {\bibfnamefont {T.}~\bibnamefont {Speck}}, \ and\
    \bibinfo {author} {\bibfnamefont {C.~P.}\ \bibnamefont {Royall}},\ }\href
    {\doibase 10.1140/epje/i2018-11662-3} {\bibfield  {journal} {\bibinfo
    {journal} {Eur. Phys. J. E}\ }\textbf {\bibinfo {volume} {41}},\ \bibinfo
    {pages} {54} (\bibinfo {year} {2018})}\BibitemShut {NoStop}%
  \bibitem [{\citenamefont {Pinchaipat}\ \emph {et~al.}(2017)\citenamefont
    {Pinchaipat}, \citenamefont {Campo}, \citenamefont {Turci}, \citenamefont
    {Hallett}, \citenamefont {Speck},\ and\ \citenamefont {Royall}}]{pinc17}%
    \BibitemOpen
    \bibfield  {author} {\bibinfo {author} {\bibfnamefont {R.}~\bibnamefont
    {Pinchaipat}}, \bibinfo {author} {\bibfnamefont {M.}~\bibnamefont {Campo}},
    \bibinfo {author} {\bibfnamefont {F.}~\bibnamefont {Turci}}, \bibinfo
    {author} {\bibfnamefont {J.~E.}\ \bibnamefont {Hallett}}, \bibinfo {author}
    {\bibfnamefont {T.}~\bibnamefont {Speck}}, \ and\ \bibinfo {author}
    {\bibfnamefont {C.~P.}\ \bibnamefont {Royall}},\ }\href {\doibase
    10.1103/PhysRevLett.119.028004} {\bibfield  {journal} {\bibinfo  {journal}
    {Phys. Rev. Lett.}\ }\textbf {\bibinfo {volume} {119}},\ \bibinfo {pages}
    {028004} (\bibinfo {year} {2017})}\BibitemShut {NoStop}%
  \bibitem [{\citenamefont {G\"otze}\ and\ \citenamefont
    {Sjogren}(1992)}]{gotz92}%
    \BibitemOpen
    \bibfield  {author} {\bibinfo {author} {\bibfnamefont {W.}~\bibnamefont
    {G\"otze}}\ and\ \bibinfo {author} {\bibfnamefont {L.}~\bibnamefont
    {Sjogren}},\ }\href {\doibase 10.1088/0034-4885/55/3/001} {\bibfield
    {journal} {\bibinfo  {journal} {Rep. Prog. Phys.}\ }\textbf {\bibinfo
    {volume} {55}},\ \bibinfo {pages} {241} (\bibinfo {year} {1992})}\BibitemShut
    {NoStop}%
  \bibitem [{\citenamefont {Keys}\ \emph {et~al.}(2011)\citenamefont {Keys},
    \citenamefont {Hedges}, \citenamefont {Garrahan}, \citenamefont {Glotzer},\
    and\ \citenamefont {Chandler}}]{keys11}%
    \BibitemOpen
    \bibfield  {author} {\bibinfo {author} {\bibfnamefont {A.~S.}\ \bibnamefont
    {Keys}}, \bibinfo {author} {\bibfnamefont {L.~O.}\ \bibnamefont {Hedges}},
    \bibinfo {author} {\bibfnamefont {J.~P.}\ \bibnamefont {Garrahan}}, \bibinfo
    {author} {\bibfnamefont {S.~C.}\ \bibnamefont {Glotzer}}, \ and\ \bibinfo
    {author} {\bibfnamefont {D.}~\bibnamefont {Chandler}},\ }\href {\doibase
    10.1103/PhysRevX.1.021013} {\bibfield  {journal} {\bibinfo  {journal} {Phys.
    Rev. X}\ }\textbf {\bibinfo {volume} {1}},\ \bibinfo {pages} {021013}
    (\bibinfo {year} {2011})}\BibitemShut {NoStop}%
  \bibitem [{\citenamefont {Cavagna}(2009)}]{cava09}%
    \BibitemOpen
    \bibfield  {author} {\bibinfo {author} {\bibfnamefont {A.}~\bibnamefont
    {Cavagna}},\ }\href {\doibase 10.1016/j.physrep.2009.03.003} {\bibfield
    {journal} {\bibinfo  {journal} {Phys. Rep.}\ }\textbf {\bibinfo {volume}
    {476}},\ \bibinfo {pages} {51} (\bibinfo {year} {2009})}\BibitemShut
    {NoStop}%
  \bibitem [{\citenamefont {Debenedetti}\ and\ \citenamefont
    {Stillinger}(2001)}]{debe01}%
    \BibitemOpen
    \bibfield  {author} {\bibinfo {author} {\bibfnamefont {P.~G.}\ \bibnamefont
    {Debenedetti}}\ and\ \bibinfo {author} {\bibfnamefont {F.~H.}\ \bibnamefont
    {Stillinger}},\ }\href {\doibase 10.1038/35065704} {\bibfield  {journal}
    {\bibinfo  {journal} {Nature}\ }\textbf {\bibinfo {volume} {410}},\ \bibinfo
    {pages} {259} (\bibinfo {year} {2001})}\BibitemShut {NoStop}%
  \bibitem [{\citenamefont {Lubchenko}\ and\ \citenamefont
    {Wolynes}(2007)}]{lubc07}%
    \BibitemOpen
    \bibfield  {author} {\bibinfo {author} {\bibfnamefont {V.}~\bibnamefont
    {Lubchenko}}\ and\ \bibinfo {author} {\bibfnamefont {P.~G.}\ \bibnamefont
    {Wolynes}},\ }\href {\doibase 10.1146/annurev.physchem.58.032806.104653}
    {\bibfield  {journal} {\bibinfo  {journal} {Annu. Rev. Phys. Chem.}\ }\textbf
    {\bibinfo {volume} {58}},\ \bibinfo {pages} {235} (\bibinfo {year}
    {2007})}\BibitemShut {NoStop}%
  \bibitem [{\citenamefont {Royall}\ and\ \citenamefont
    {Williams}(2015)}]{roya15}%
    \BibitemOpen
    \bibfield  {author} {\bibinfo {author} {\bibfnamefont {C.~P.}\ \bibnamefont
    {Royall}}\ and\ \bibinfo {author} {\bibfnamefont {S.~R.}\ \bibnamefont
    {Williams}},\ }\href {\doibase 10.1016/j.physrep.2014.11.004} {\bibfield
    {journal} {\bibinfo  {journal} {Phys. Rep.}\ }\textbf {\bibinfo {volume}
    {560}},\ \bibinfo {pages} {1} (\bibinfo {year} {2015})}\BibitemShut {NoStop}%
  \bibitem [{\citenamefont {Adam}\ and\ \citenamefont {Gibbs}(1965)}]{adam65}%
    \BibitemOpen
    \bibfield  {author} {\bibinfo {author} {\bibfnamefont {G.}~\bibnamefont
    {Adam}}\ and\ \bibinfo {author} {\bibfnamefont {J.~H.}\ \bibnamefont
    {Gibbs}},\ }\href {\doibase 10.1063/1.1696442} {\bibfield  {journal}
    {\bibinfo  {journal} {J. Chem. Phys.}\ }\textbf {\bibinfo {volume} {43}},\
    \bibinfo {pages} {139} (\bibinfo {year} {1965})}\BibitemShut {NoStop}%
  \bibitem [{\citenamefont {Kirkpatrick}\ \emph {et~al.}(1989)\citenamefont
    {Kirkpatrick}, \citenamefont {Thirumalai},\ and\ \citenamefont
    {Wolynes}}]{kirk89}%
    \BibitemOpen
    \bibfield  {author} {\bibinfo {author} {\bibfnamefont {T.~R.}\ \bibnamefont
    {Kirkpatrick}}, \bibinfo {author} {\bibfnamefont {D.}~\bibnamefont
    {Thirumalai}}, \ and\ \bibinfo {author} {\bibfnamefont {P.~G.}\ \bibnamefont
    {Wolynes}},\ }\href {\doibase 10.1103/PhysRevA.40.1045} {\bibfield  {journal}
    {\bibinfo  {journal} {Phys. Rev. A}\ }\textbf {\bibinfo {volume} {40}},\
    \bibinfo {pages} {1045} (\bibinfo {year} {1989})}\BibitemShut {NoStop}%
  \bibitem [{\citenamefont {Bouchaud}\ and\ \citenamefont
    {Biroli}(2004)}]{bouchaud04}%
    \BibitemOpen
    \bibfield  {author} {\bibinfo {author} {\bibfnamefont {J.-P.}\ \bibnamefont
    {Bouchaud}}\ and\ \bibinfo {author} {\bibfnamefont {G.}~\bibnamefont
    {Biroli}},\ }\href {\doibase 10.1063/1.1796231} {\bibfield  {journal}
    {\bibinfo  {journal} {J. Chem. Phys.}\ }\textbf {\bibinfo {volume} {121}},\
    \bibinfo {pages} {7347} (\bibinfo {year} {2004})}\BibitemShut {NoStop}%
  \bibitem [{\citenamefont {Palmer}\ \emph {et~al.}(1984)\citenamefont {Palmer},
    \citenamefont {Stein}, \citenamefont {Abrahams},\ and\ \citenamefont
    {Anderson}}]{palm84}%
    \BibitemOpen
    \bibfield  {author} {\bibinfo {author} {\bibfnamefont {R.~G.}\ \bibnamefont
    {Palmer}}, \bibinfo {author} {\bibfnamefont {D.~L.}\ \bibnamefont {Stein}},
    \bibinfo {author} {\bibfnamefont {E.}~\bibnamefont {Abrahams}}, \ and\
    \bibinfo {author} {\bibfnamefont {P.~W.}\ \bibnamefont {Anderson}},\ }\href
    {\doibase 10.1103/PhysRevLett.53.958} {\bibfield  {journal} {\bibinfo
    {journal} {Phys. Rev. Lett.}\ }\textbf {\bibinfo {volume} {53}},\ \bibinfo
    {pages} {958} (\bibinfo {year} {1984})}\BibitemShut {NoStop}%
  \bibitem [{\citenamefont {Elmatad}\ \emph {et~al.}(2009)\citenamefont
    {Elmatad}, \citenamefont {Chandler},\ and\ \citenamefont
    {Garrahan}}]{elma09}%
    \BibitemOpen
    \bibfield  {author} {\bibinfo {author} {\bibfnamefont {Y.~S.}\ \bibnamefont
    {Elmatad}}, \bibinfo {author} {\bibfnamefont {D.}~\bibnamefont {Chandler}}, \
    and\ \bibinfo {author} {\bibfnamefont {J.~P.}\ \bibnamefont {Garrahan}},\
    }\href {\doibase 10.1021/jp810362g} {\bibfield  {journal} {\bibinfo
    {journal} {J. Phys. Chem. B}\ }\textbf {\bibinfo {volume} {113}},\ \bibinfo
    {pages} {5563} (\bibinfo {year} {2009})}\BibitemShut {NoStop}%
  \bibitem [{\citenamefont {Honeycutt}\ and\ \citenamefont
    {Andersen}(1987)}]{hone87}%
    \BibitemOpen
    \bibfield  {author} {\bibinfo {author} {\bibfnamefont {J.~D.}\ \bibnamefont
    {Honeycutt}}\ and\ \bibinfo {author} {\bibfnamefont {H.~C.}\ \bibnamefont
    {Andersen}},\ }\href {\doibase 10.1021/j100303a014} {\bibfield  {journal}
    {\bibinfo  {journal} {J. Phys. Chem.}\ }\textbf {\bibinfo {volume} {91}},\
    \bibinfo {pages} {4950} (\bibinfo {year} {1987})}\BibitemShut {NoStop}%
  \bibitem [{\citenamefont {Coslovich}(2011)}]{cosl11}%
    \BibitemOpen
    \bibfield  {author} {\bibinfo {author} {\bibfnamefont {D.}~\bibnamefont
    {Coslovich}},\ }\href {\doibase 10.1103/PhysRevE.83.051505} {\bibfield
    {journal} {\bibinfo  {journal} {Phys. Rev. E}\ }\textbf {\bibinfo {volume}
    {83}},\ \bibinfo {pages} {051505} (\bibinfo {year} {2011})}\BibitemShut
    {NoStop}%
  \bibitem [{\citenamefont {Malins}\ \emph {et~al.}(2013)\citenamefont {Malins},
    \citenamefont {Williams}, \citenamefont {Eggers},\ and\ \citenamefont
    {Royall}}]{mali13}%
    \BibitemOpen
    \bibfield  {author} {\bibinfo {author} {\bibfnamefont {A.}~\bibnamefont
    {Malins}}, \bibinfo {author} {\bibfnamefont {S.~R.}\ \bibnamefont
    {Williams}}, \bibinfo {author} {\bibfnamefont {J.}~\bibnamefont {Eggers}}, \
    and\ \bibinfo {author} {\bibfnamefont {C.~P.}\ \bibnamefont {Royall}},\
    }\href {\doibase http://dx.doi.org/10.1063/1.4832897} {\bibfield  {journal}
    {\bibinfo  {journal} {J. Chem. Phys.}\ }\textbf {\bibinfo {volume} {139}},\
    \bibinfo {pages} {234506} (\bibinfo {year} {2013})}\BibitemShut {NoStop}%
  \bibitem [{\citenamefont {Speck}\ \emph {et~al.}(2012)\citenamefont {Speck},
    \citenamefont {Malins},\ and\ \citenamefont {Royall}}]{spec12b}%
    \BibitemOpen
    \bibfield  {author} {\bibinfo {author} {\bibfnamefont {T.}~\bibnamefont
    {Speck}}, \bibinfo {author} {\bibfnamefont {A.}~\bibnamefont {Malins}}, \
    and\ \bibinfo {author} {\bibfnamefont {C.~P.}\ \bibnamefont {Royall}},\
    }\href {\doibase 10.1103/PhysRevLett.109.195703} {\bibfield  {journal}
    {\bibinfo  {journal} {Phys. Rev. Lett.}\ }\textbf {\bibinfo {volume} {109}},\
    \bibinfo {pages} {195703} (\bibinfo {year} {2012})}\BibitemShut {NoStop}%
  \bibitem [{\citenamefont {Turci}\ \emph {et~al.}(2017)\citenamefont {Turci},
    \citenamefont {Royall},\ and\ \citenamefont {Speck}}]{turci17}%
    \BibitemOpen
    \bibfield  {author} {\bibinfo {author} {\bibfnamefont {F.}~\bibnamefont
    {Turci}}, \bibinfo {author} {\bibfnamefont {C.~P.}\ \bibnamefont {Royall}}, \
    and\ \bibinfo {author} {\bibfnamefont {T.}~\bibnamefont {Speck}},\ }\href
    {\doibase 10.1103/physrevx.7.031028} {\bibfield  {journal} {\bibinfo
    {journal} {Phys. Rev. X}\ }\textbf {\bibinfo {volume} {7}},\ \bibinfo {pages}
    {031028} (\bibinfo {year} {2017})}\BibitemShut {NoStop}%
  \bibitem [{\citenamefont {Hocky}\ \emph {et~al.}(2014)\citenamefont {Hocky},
    \citenamefont {Coslovich}, \citenamefont {Ikeda},\ and\ \citenamefont
    {Reichman}}]{hock14}%
    \BibitemOpen
    \bibfield  {author} {\bibinfo {author} {\bibfnamefont {G.~M.}\ \bibnamefont
    {Hocky}}, \bibinfo {author} {\bibfnamefont {D.}~\bibnamefont {Coslovich}},
    \bibinfo {author} {\bibfnamefont {A.}~\bibnamefont {Ikeda}}, \ and\ \bibinfo
    {author} {\bibfnamefont {D.~R.}\ \bibnamefont {Reichman}},\ }\href {\doibase
    10.1103/PhysRevLett.113.157801} {\bibfield  {journal} {\bibinfo  {journal}
    {Phys. Rev. Lett.}\ }\textbf {\bibinfo {volume} {113}},\ \bibinfo {pages}
    {157801} (\bibinfo {year} {2014})}\BibitemShut {NoStop}%
  \bibitem [{\citenamefont {Coslovich}\ and\ \citenamefont
    {Jack}(2016)}]{cosl16}%
    \BibitemOpen
    \bibfield  {author} {\bibinfo {author} {\bibfnamefont {D.}~\bibnamefont
    {Coslovich}}\ and\ \bibinfo {author} {\bibfnamefont {R.~L.}\ \bibnamefont
    {Jack}},\ }\href {\doibase 10.1088/1742-5468/2016/07/074012} {\bibfield
    {journal} {\bibinfo  {journal} {J. Stat. Mech. Theory Exp.}\ }\textbf
    {\bibinfo {volume} {2016}},\ \bibinfo {pages} {074012} (\bibinfo {year}
    {2016})}\BibitemShut {NoStop}%
  \bibitem [{\citenamefont {Schoenholz}\ \emph {et~al.}(2016)\citenamefont
    {Schoenholz}, \citenamefont {Cubuk}, \citenamefont {Sussman}, \citenamefont
    {Kaxiras},\ and\ \citenamefont {Liu}}]{scho16}%
    \BibitemOpen
    \bibfield  {author} {\bibinfo {author} {\bibfnamefont {S.~S.}\ \bibnamefont
    {Schoenholz}}, \bibinfo {author} {\bibfnamefont {E.~D.}\ \bibnamefont
    {Cubuk}}, \bibinfo {author} {\bibfnamefont {D.~M.}\ \bibnamefont {Sussman}},
    \bibinfo {author} {\bibfnamefont {E.}~\bibnamefont {Kaxiras}}, \ and\
    \bibinfo {author} {\bibfnamefont {A.~J.}\ \bibnamefont {Liu}},\ }\href
    {\doibase 10.1038/nphys3644} {\bibfield  {journal} {\bibinfo  {journal} {Nat.
    Phys.}\ }\textbf {\bibinfo {volume} {12}},\ \bibinfo {pages} {469} (\bibinfo
    {year} {2016})}\BibitemShut {NoStop}%
  \bibitem [{\citenamefont {Kauzmann}(1948)}]{kauz48}%
    \BibitemOpen
    \bibfield  {author} {\bibinfo {author} {\bibfnamefont {W.}~\bibnamefont
    {Kauzmann}},\ }\href {\doibase 10.1021/cr60135a002} {\bibfield  {journal}
    {\bibinfo  {journal} {Chem. Rev.}\ }\textbf {\bibinfo {volume} {43}},\
    \bibinfo {pages} {219} (\bibinfo {year} {1948})}\BibitemShut {NoStop}%
  \bibitem [{\citenamefont {Stillinger}(1988)}]{stil88}%
    \BibitemOpen
    \bibfield  {author} {\bibinfo {author} {\bibfnamefont {F.~H.}\ \bibnamefont
    {Stillinger}},\ }\href {\doibase http://dx.doi.org/10.1063/1.454295}
    {\bibfield  {journal} {\bibinfo  {journal} {J. Chem. Phys.}\ }\textbf
    {\bibinfo {volume} {88}},\ \bibinfo {pages} {7818} (\bibinfo {year}
    {1988})}\BibitemShut {NoStop}%
  \bibitem [{\citenamefont {Moynihan}\ and\ \citenamefont
    {Angell}(2000)}]{moyn00}%
    \BibitemOpen
    \bibfield  {author} {\bibinfo {author} {\bibfnamefont {C.~T.}\ \bibnamefont
    {Moynihan}}\ and\ \bibinfo {author} {\bibfnamefont {C.~A.}\ \bibnamefont
    {Angell}},\ }\href {\doibase 10.1016/S0022-3093(00)00198-8} {\bibfield
    {journal} {\bibinfo  {journal} {J. Non-Cryst. Solids}\ }\textbf {\bibinfo
    {volume} {274}},\ \bibinfo {pages} {131} (\bibinfo {year}
    {2000})}\BibitemShut {NoStop}%
  \bibitem [{\citenamefont {Biroli}\ \emph {et~al.}(2005)\citenamefont {Biroli},
    \citenamefont {Bouchaud},\ and\ \citenamefont {Tarjus}}]{biro05}%
    \BibitemOpen
    \bibfield  {author} {\bibinfo {author} {\bibfnamefont {G.}~\bibnamefont
    {Biroli}}, \bibinfo {author} {\bibfnamefont {J.-P.}\ \bibnamefont
    {Bouchaud}}, \ and\ \bibinfo {author} {\bibfnamefont {G.}~\bibnamefont
    {Tarjus}},\ }\href {\doibase 10.1063/1.1955527} {\bibfield  {journal}
    {\bibinfo  {journal} {J. Chem. Phys.}\ }\textbf {\bibinfo {volume} {123}},\
    \bibinfo {pages} {044510} (\bibinfo {year} {2005})}\BibitemShut {NoStop}%
  \bibitem [{\citenamefont {Chandler}\ and\ \citenamefont
    {Garrahan}(2005)}]{chan05}%
    \BibitemOpen
    \bibfield  {author} {\bibinfo {author} {\bibfnamefont {D.}~\bibnamefont
    {Chandler}}\ and\ \bibinfo {author} {\bibfnamefont {J.~P.}\ \bibnamefont
    {Garrahan}},\ }\href {\doibase 10.1063/1.1955528} {\bibfield  {journal}
    {\bibinfo  {journal} {J. Chem. Phys.}\ }\textbf {\bibinfo {volume} {123}},\
    \bibinfo {pages} {044511} (\bibinfo {year} {2005})}\BibitemShut {NoStop}%
  \bibitem [{\citenamefont {Cammarota}\ and\ \citenamefont
    {Biroli}(2012)}]{camm12}%
    \BibitemOpen
    \bibfield  {author} {\bibinfo {author} {\bibfnamefont {C.}~\bibnamefont
    {Cammarota}}\ and\ \bibinfo {author} {\bibfnamefont {G.}~\bibnamefont
    {Biroli}},\ }\href {\doibase 10.1073/pnas.1111582109} {\bibfield  {journal}
    {\bibinfo  {journal} {Proc. Natl. Acad. Sci. U.S.A.}\ }\textbf {\bibinfo
    {volume} {109}},\ \bibinfo {pages} {8850} (\bibinfo {year}
    {2012})}\BibitemShut {NoStop}%
  \bibitem [{\citenamefont {Gokhale}\ \emph {et~al.}(2014)\citenamefont
    {Gokhale}, \citenamefont {Nagamanasa}, \citenamefont {Ganapathy},\ and\
    \citenamefont {Sood}}]{gokh14}%
    \BibitemOpen
    \bibfield  {author} {\bibinfo {author} {\bibfnamefont {S.}~\bibnamefont
    {Gokhale}}, \bibinfo {author} {\bibfnamefont {K.~H.}\ \bibnamefont
    {Nagamanasa}}, \bibinfo {author} {\bibfnamefont {R.}~\bibnamefont
    {Ganapathy}}, \ and\ \bibinfo {author} {\bibfnamefont {A.~K.}\ \bibnamefont
    {Sood}},\ }\href {\doibase 10.1038/ncomms5685} {\bibfield  {journal}
    {\bibinfo  {journal} {Nat. Commun.}\ }\textbf {\bibinfo {volume} {5}},\
    \bibinfo {pages} {4685} (\bibinfo {year} {2014})}\BibitemShut {NoStop}%
  \bibitem [{\citenamefont {Albert}\ \emph {et~al.}(2016)\citenamefont {Albert},
    \citenamefont {Bauer}, \citenamefont {Michl}, \citenamefont {Biroli},
    \citenamefont {Bouchaud}, \citenamefont {Loidl}, \citenamefont
    {Lunkenheimer}, \citenamefont {Tourbot}, \citenamefont {Wiertel-Gasquet},\
    and\ \citenamefont {Ladieu}}]{albe16}%
    \BibitemOpen
    \bibfield  {author} {\bibinfo {author} {\bibfnamefont {S.}~\bibnamefont
    {Albert}}, \bibinfo {author} {\bibfnamefont {T.}~\bibnamefont {Bauer}},
    \bibinfo {author} {\bibfnamefont {M.}~\bibnamefont {Michl}}, \bibinfo
    {author} {\bibfnamefont {G.}~\bibnamefont {Biroli}}, \bibinfo {author}
    {\bibfnamefont {J.-P.}\ \bibnamefont {Bouchaud}}, \bibinfo {author}
    {\bibfnamefont {A.}~\bibnamefont {Loidl}}, \bibinfo {author} {\bibfnamefont
    {P.}~\bibnamefont {Lunkenheimer}}, \bibinfo {author} {\bibfnamefont
    {R.}~\bibnamefont {Tourbot}}, \bibinfo {author} {\bibfnamefont
    {C.}~\bibnamefont {Wiertel-Gasquet}}, \ and\ \bibinfo {author} {\bibfnamefont
    {F.}~\bibnamefont {Ladieu}},\ }\href {\doibase 10.1126/science.aaf3182}
    {\bibfield  {journal} {\bibinfo  {journal} {Science}\ }\textbf {\bibinfo
    {volume} {352}},\ \bibinfo {pages} {1308} (\bibinfo {year}
    {2016})}\BibitemShut {NoStop}%
  \bibitem [{\citenamefont {Jung}\ \emph {et~al.}(2004)\citenamefont {Jung},
    \citenamefont {Garrahan},\ and\ \citenamefont {Chandler}}]{jung04}%
    \BibitemOpen
    \bibfield  {author} {\bibinfo {author} {\bibfnamefont {Y.}~\bibnamefont
    {Jung}}, \bibinfo {author} {\bibfnamefont {J.~P.}\ \bibnamefont {Garrahan}},
    \ and\ \bibinfo {author} {\bibfnamefont {D.}~\bibnamefont {Chandler}},\
    }\href {\doibase 10.1103/PhysRevE.69.061205} {\bibfield  {journal} {\bibinfo
    {journal} {Phys. Rev. E}\ }\textbf {\bibinfo {volume} {69}},\ \bibinfo
    {pages} {061205} (\bibinfo {year} {2004})}\BibitemShut {NoStop}%
  \bibitem [{\citenamefont {Keys}\ \emph {et~al.}(2013)\citenamefont {Keys},
    \citenamefont {Garrahan},\ and\ \citenamefont {Chandler}}]{keys13a}%
    \BibitemOpen
    \bibfield  {author} {\bibinfo {author} {\bibfnamefont {A.~S.}\ \bibnamefont
    {Keys}}, \bibinfo {author} {\bibfnamefont {J.~P.}\ \bibnamefont {Garrahan}},
    \ and\ \bibinfo {author} {\bibfnamefont {D.}~\bibnamefont {Chandler}},\
    }\href {\doibase 10.1073/pnas.1302665110} {\bibfield  {journal} {\bibinfo
    {journal} {Proc. Natl. Acad. Sci. U.S.A.}\ }\textbf {\bibinfo {volume}
    {110}},\ \bibinfo {pages} {4482} (\bibinfo {year} {2013})}\BibitemShut
    {NoStop}%
  \end{thebibliography}
\end{document}